\documentclass[12pt]{article}
\usepackage[dvips]{graphicx}

\usepackage{color,graphics}
\definecolor{orange}{rgb}{.95,.75,0}

\begin{document}
\def\revise{\color{blue}}

\title{Translating oscillatory nonlinear structure in a plasma boundary}
\author{F. Haas and P. K. Shukla}
\date{\relax}
\maketitle
\begin{center}
{Institut f\"ur Theoretische Physik IV \\ 
Ruhr-Universit\"at Bochum\\ D-44780 Bochum Germany}
\end{center}

\begin{abstract}
\noindent
By means of a Madelung decomposition, exact periodic traveling solutions are constructed for a modified nonlinear Schr\"odinger equation derived by Stenflo and Gradov, describing electrostatic surface waves in semi-infinite plasma. The condition for the existence of bi-stable equilibria is discussed.  A conservation law as well as the modulational instability admitted by the model are analyzed. 
\end{abstract}

\hspace{.25cm} PACS: 52.35.Mw, 52.35.Fp, 41.20.Cv

\section{Introduction}
The description of nonlinear plasma surface waves is a growing subfield in plasma physics \cite{Stenflo1, Brodin}. Reasons for this research can be found in industrial applications \cite{Stenflo2}, plasma microwave diagnostics \cite{Shivarova}, some aspects in laser fusion \cite{Yu} and space plasma physics \cite{Buti}. In particular, much attention has been paid to electrostatic soliton-like oscillations, or oscillons \cite{Stenflo3, Stenflo4}, for cylindrical \cite{Stenflo6} as well as spherical \cite{Chakra, Stenflo7} plasmas. Similar oscillon solutions have been also reported as highly localized ring structures in Bose-Einstein condensates \cite{Baizakov}. The present work, on the other hand, consider {\it one-dimensional} nonlinear periodic structures in a plasma interface. These oscillatory solutions are found through a different treatment of a modified nonlinear Schr\"odinger equation derived earlier by Stenflo and Gradov \cite{Stenflo, Gradov}. This equation resembles some classes of PDEs discussed in the context of dissipative quantum mechanics \cite{Doebner1}-\cite{Doebner3} and can be shown to admit exact solitary solutions \cite{Gradov2, Malomed}.    

This work is organized as follows. In Section II the generalized nonlinear Schr\"odinger equation for surface waves in a sharp plasma interface is reviewed. Instead of removing the time-derivative term as in previous contributions \cite{Stenflo}, here a Madelung transformation is applied, so that a system of partial differential equations for the amplitude and the phase of the envelope electrostatic field is derived. In Section III, the problem of finding traveling solutions for this set of partial differential equations is cast in the form of an autonomous Hamiltonian system.  Some basic properties of the model equation are considered in Section IV, including the study of its modulational instability and the derivation of a conservation law. The shape of the associated Sagdeev potential and the different possible equilibria are analyzed. Illustrative trajectories are computed numerically. Section V is reserved to the conclusions.  

\section{Madelung transformation}
Our starting point is Eq. (5) of Ref. \cite{Stenflo}, specialized to the case of a plane boundary ($z = 0$),
\begin{eqnarray}
\frac{8\,ik^2}{\omega}\,\partial_{t}\phi_0 + \partial^{2}_{x}\phi_0 - \frac{(\partial_{x}\phi_0)^2}{\phi_0} + 2\,\beta\gamma\,k|\phi_0|\,\phi_0 + \beta^2 |\phi_0|^2 \phi_0 + 2\,k^2\Delta\,\phi_0 = 0 \,. \nonumber \\ 
\label{e1}
\end{eqnarray}
Formerly \cite{Gradov}, a similar equation was considered, but without the time-de\-ri\-va\-ti\-ve term, as well as the terms proportional to $\gamma$  (defined below Eq. (\ref{e4})) and $\Delta = 2 - \omega_{p}^2/\omega^2$, a measure of the mismatch to the linear surface plasmon frequency $\omega_{p}/\sqrt{2}$, where $\omega_p$ is the plasma frequency. The purpose of the present communication is to show that a plasma surface can admit exact one-dimensional oscillatory structures, provided these contributions are taken into account. In the former case, an exact localized, surface solitary wave was found \cite{Gradov}. 

Let us briefly examine the meaning of the different terms in Eq. (\ref{e1}), as well as its derivation \cite{Stenflo, Gradov}. Consider a plasma boundary at $z = 0$, with a fixed ionic background of constant density for $z > 0$ and vacuum for $z < 0$, and electrostatic oscillations. Then the scalar potential $\phi(x,z,t)\,\exp[i(k x-\omega\,t)]$ satisfy the Laplace equation for $z < 0$, while inside the plasma it satisfies $\nabla\cdot\,(\epsilon\nabla\phi) = 0$. Here the dielectric function $\epsilon$ is 
\begin{equation}
\label{e2}
\epsilon = 1 - \frac{\omega_{p}^2}{\omega^2} + \frac{2i\omega_{p}^2}{\omega^3}\,\partial_{t}\ln\phi - \frac{\beta^2}{2k^2}|\phi|^2 \,.
\end{equation}
In Eq. (2), the $\sim \beta^2$ term is a small polynomial nonlinear correction. We consider the case of cold electrons and immobile ions, so that $\beta = - 4 e k^3/m\omega^2$, where $-e$ and $m$ are the electron charge and mass \cite{Gradov}. Further, assume 
\begin{equation}
\label{e3}
\phi = \phi_{0}(x,t)\exp\left[-k\,z + \int_{0}^{z}dz'\,k_{z}(x,z')\right] \,,
\end{equation}
for $z > 0$, where $\phi_0$ is slowly varying both in space and time. Matching Eq. (\ref{e3}) to the appropriated analytic continuation solution outside the plasma, taking into account the continuity of $\phi$ and the normal component of the electric field at $z = 0$, one derive Eq. (\ref{e1}). For this purpose, one uses the first order Taylor expansion of $k_z \ll k$ in powers of $\phi$ and 
\begin{equation}
\label{e4}
\left(\frac{\partial k_z}{\partial z}\right)_{z=0} \simeq 2\,\beta\gamma\,k|\phi| = - \frac{8\gamma ek^4}{m\omega^2}\,|\phi| \,,
\end{equation}
which implicitly define the phenomenological non-dimensional parameter $\gamma$. See Refs. \cite{Stenflo, Gradov} for more details.

In Ref. \cite{Stenflo}, instead of the plane $xy$ the boundary of the plasma was taken as deformable due to the ponderomotive force. The plasma interface was parametrized by a suitable function $z = \xi(x,t)$, allowing to eliminate the time-derivative term in Eq. (\ref{e1}) in a moving frame. Here, we follow a different approach, taking a fixed $z = 0$ plasma interface and observing that Eq. (\ref{e1}) is a modified nonlinear Schr\"odinger equation. This suggest the application of a Madelung, or eikonal, decomposition,
\begin{equation}
\label{e5}
\phi_0 = A\exp(iS) \,,
\end{equation}
where the amplitude $A = A(x,t)$ and the phase $S = S(x,t)$ are real functions. Inserting Eq. (\ref{e5}) into Eq. (\ref{e1}) and separating the real and imaginary parts, it follows that
\begin{eqnarray}
\label{e6}
\partial^{2}_{x}S &=& - \frac{8\,k^2}{\omega}\frac{\partial_{t}A}{A} \,,\\
\label{e7}
\frac{8\,k^2}{\omega}\partial_{t}S &=& \frac{\partial^{2}_{x}A}{A} - \frac{[\partial_{x}A]^2}{A^2} + 2\beta\gamma\, k\,A + \beta^2 A^2 + 2k^2\Delta \,.
\end{eqnarray}
Eliminating $S$ between Eqs. (\ref{e6}-\ref{e7}), it follows that
\begin{equation}
\label{e8}
\partial_{x}^2\left(\partial_{x}^2\psi + 2\beta\gamma\,k\,A_0 e^{\psi}+\beta^2 A_{0}^2\,e^{2\psi}\right) + \left(8\,k^2/\omega\right)^2\partial_{t}^2\psi = 0 \,,
\end{equation}
where we have defined
\begin{equation}
\label{e9}
\psi = \ln(A/A_{0}) \,,
\end{equation}
in terms of a reference amplitude $A_0$.

\section{Sagdeev potential}
Assume a traveling wave solution $\psi = \psi(x-ut)$ for Eq. (\ref{e8}), where $u$ is a constant parameter. Using decaying boundary conditions, Eq. (\ref{e8}) is converted into the autonomous one-dimensional Hamiltonian system $d^{2}\psi/dX^2 = - dV/d\psi$, where $X = x-ut$ and $V = V(\psi)$ is the Sagdeev potential 
\begin{equation}
\label{e10}
V = 32\left(\frac{k^2 u}{\omega}\right)^2\psi^2 + 2\beta\gamma\,k\,A_0 \,(e^\psi - \psi-1) + \frac{\beta^2 A_{0}^2}{2}\,(e^{2\psi}-2\psi-1) \,.
\end{equation}
Hence evidently the problem is reducible to a quadrature using the energy integral $H = (\psi')^2/2 + V(\psi)$. Notice that the above Sagdeev potential is different from the one derived for similar nonlinear surface plasma wave problems \cite{Stenflo7}, \cite{Stenflo8, Lan}, due to the different starting assumptions and modeling. 

To complete the solution, Eqs. (\ref{e6}-\ref{e7}) can be used to obtain the phase $S = S(X)$ through 
\begin{equation}
\label{e11}
\frac{dS}{dX} = \frac{8\,k^2\,u\,\psi}{\omega} - \frac{\omega}{8\,k^2u}\,(2\beta\gamma k\,A_0 \,+\beta^2A_{0}^2+2k^2\Delta) \,.
\end{equation}

\begin{figure}
\begin{center}
\includegraphics{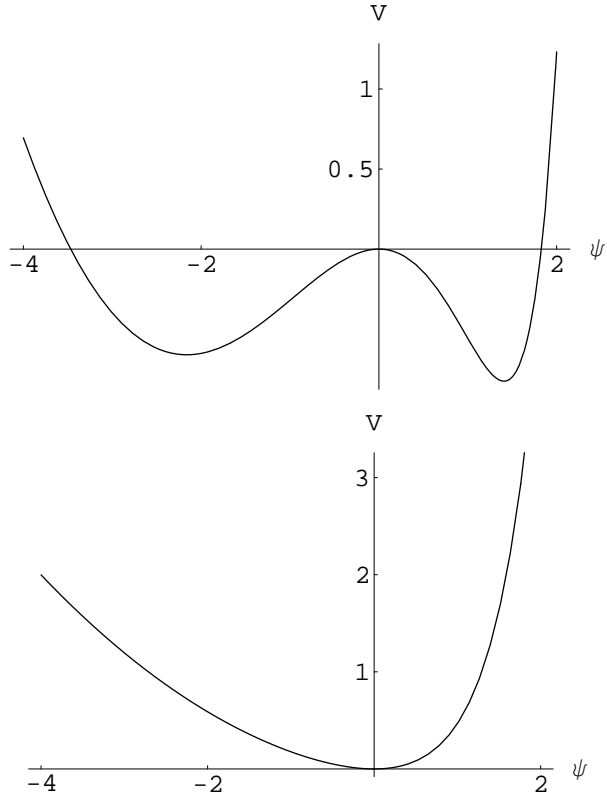}
\caption{(Upper) Sagdeev potential in Eq. (\ref{e10}) for $32 (k^2u/\omega)^2 = 1/2, -2\beta\gamma\,k|A_0| = 3, \beta^2 A_{0}^2/2 = 1/4$; (Bottom) Sagdeev potential in Eq. (\ref{e10}) for $32 (k^2u/\omega)^2 = -2\beta\gamma\,k|A_0| = \beta^2 A_{0}^2/2 = 1/10$.}
\label{fig1}
\end{center}
\end{figure}

\begin{figure}
\begin{center}
\includegraphics{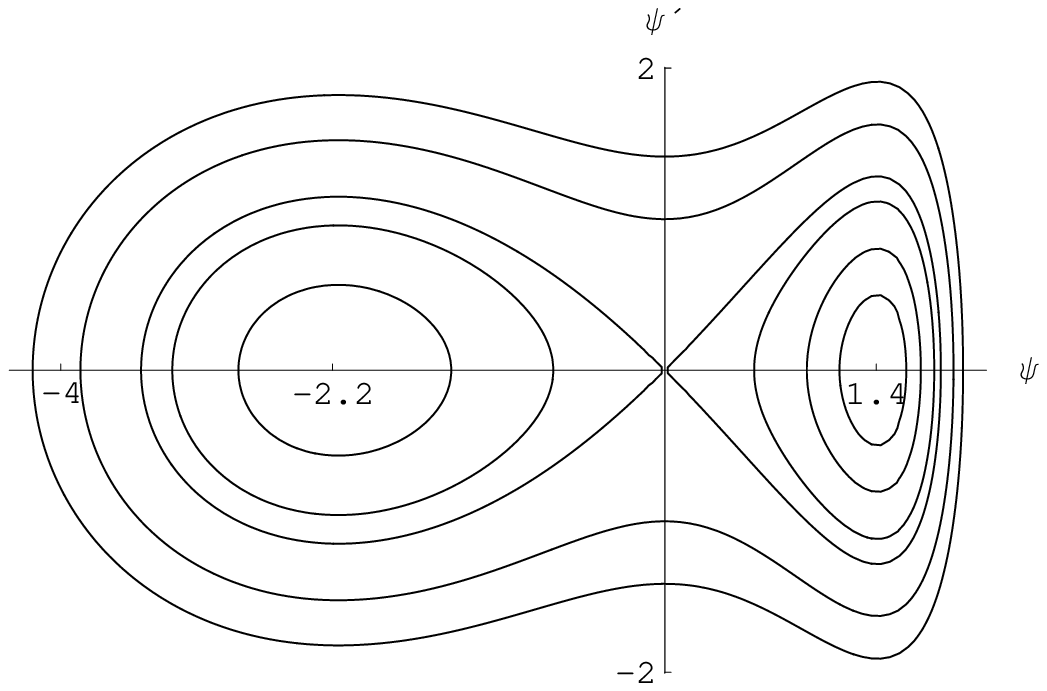}
\caption{Phase-space curves in a typical bi-stable equilibrium case of the Sagdeev potential in Eq. (\ref{e10}). Parameters: $32 (k^2u/\omega)^2 = 1/2, -2\beta\gamma\,k|A_0| = 3, \beta^2 A_{0}^2/2 = 1/4$.}
\label{fig2}
\end{center}
\end{figure}

Because the envelope $\phi_0$ is slowly varying, necessarily $u$ is a small quantity. However, we exclude the case where it is strictly zero. In this way,  $V \sim \psi^2$ as $\psi \rightarrow - \infty$. Since $V \sim e^{2\psi}$ as $\psi \rightarrow \infty$, we then have an asymmetric trap sustaining periodic nonlinear oscillations of $\psi$. Moreover, observing that $V(0) = V'(0) = 0$, we conclude that a double well is present whenever $V''(0) < 0$, see Fig. (\ref{fig1}). In the present work we assume that the condition $V''(0) < 0$, or 
\begin{equation}
\label{e12}
-\beta\gamma k\,A_0 > 32\left(\frac{k^2 u}{\omega}\right)^2 + \beta^2A_{0}^2 
\end{equation}
is fulfilled, so that bi-stability is possible for a negative pseudo-energy $H$. The corresponding phase-space curves are in Fig. (\ref{fig2}). 

Since $\beta\,k$ is a negative quantity, to satisfy the inequality (\ref{e12}) a necessary condition is that $\gamma > 0$. In view of Eqs. (\ref{e3}-\ref{e4}), this is a most natural requirement, since we expect a bounded scalar potential. Moreover, approximating $\omega$ by its linear expression $\omega_p/\sqrt{2}$ and using the definition of $\beta$, it is possible to rewrite Eq. (\ref{e12}) as 
\begin{equation}
\label{e13}
\left(\frac{e\,A_0\,k}{m\,\omega_p}\right)^2 < \frac{\gamma e\,A_0 \,}{8\,m} - u^2 \,,
\end{equation}
which impose a lower bound on the wavelengths for bi-stable oscillations. In addition, the velocity $u$ should not be too large. 

\begin{figure}
\begin{center}
\includegraphics{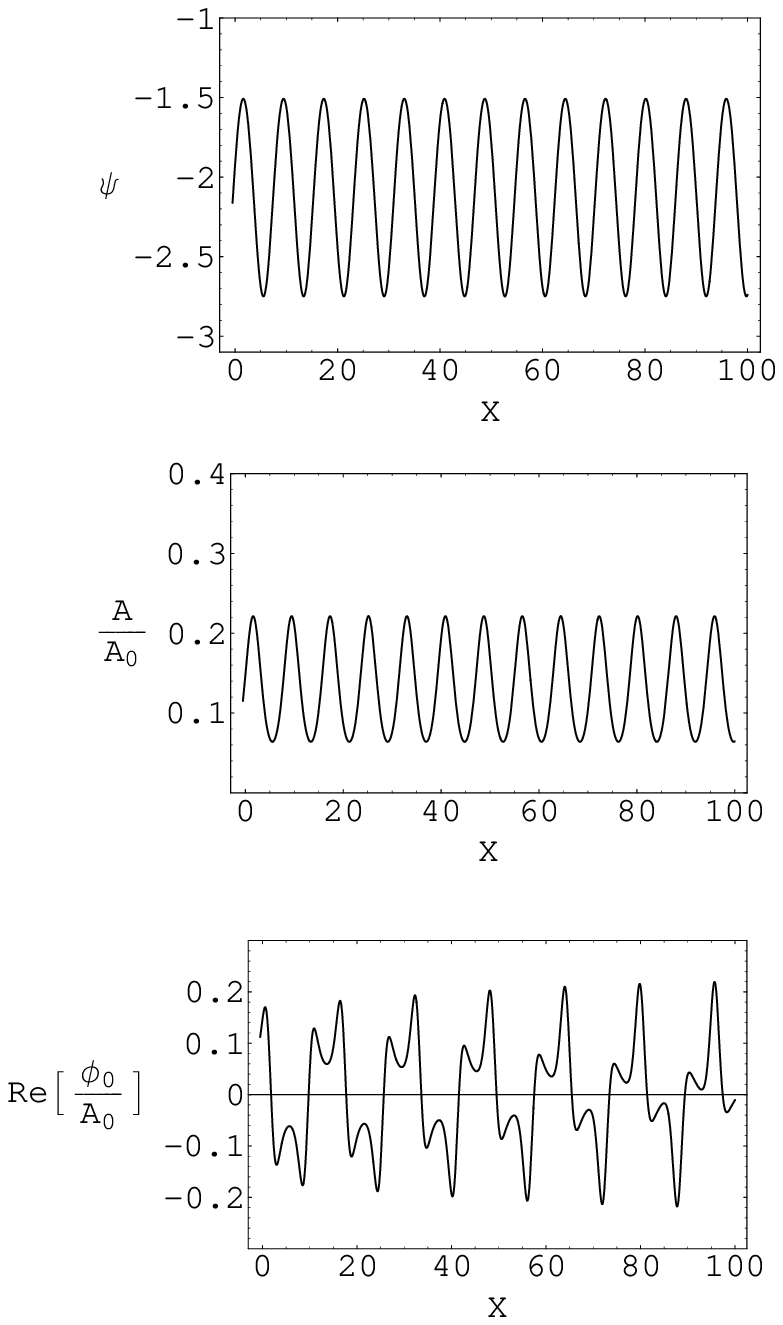}
\caption{Auxiliary function $\psi$ (on the top), amplitude $A$ (on the mid) and real part of the envelope $\phi_0$ (on the bottom) for linear oscillations near the fixed point on the left for the Sagdeev potential in Eq. (\ref{e10}). Parameters: $32 (k^2u/\omega)^2 = 1/2, -2\beta\gamma\,k|A_0| = 3, \beta^2 A_{0}^2/2 = 1/4, \Delta = 0$. Initial conditions: $\psi(0) = -2.2, \psi'(0) = 0.5$.}
\label{fig3}
\end{center}
\end{figure}

\begin{figure}
\begin{center}
\includegraphics{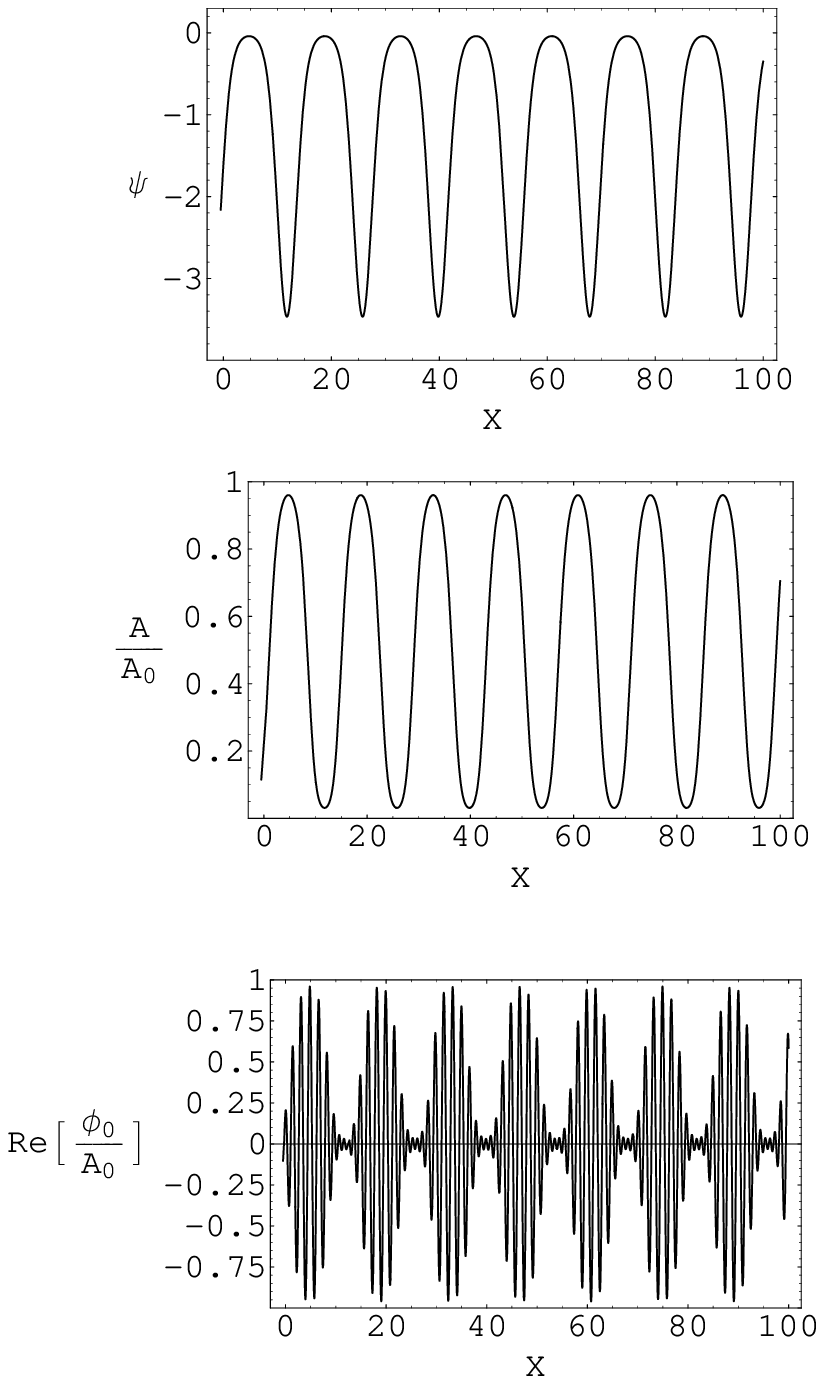}
\caption{Auxiliary function $\psi$ (on the top), amplitude $A$ (on the mid) and real part of the envelope $\phi_0$ (on the bottom) for nonlinear oscillations around the fixed point on the left for the Sagdeev potential in Eq. (\ref{e10}). Parameters: $32 (k^2u/\omega)^2 = 1/2, -2\beta\gamma\,k|A_0| = 3, \beta^2 A_{0}^2/2 = 1/4, \Delta = 0$. Initial conditions: $\psi(0) = -2.2, \psi'(0) = 1.2$. The pseudo-energy is then $H = - 10^{-3}$.}
\label{fig4}
\end{center}
\end{figure}

\begin{figure}
\begin{center}
\includegraphics{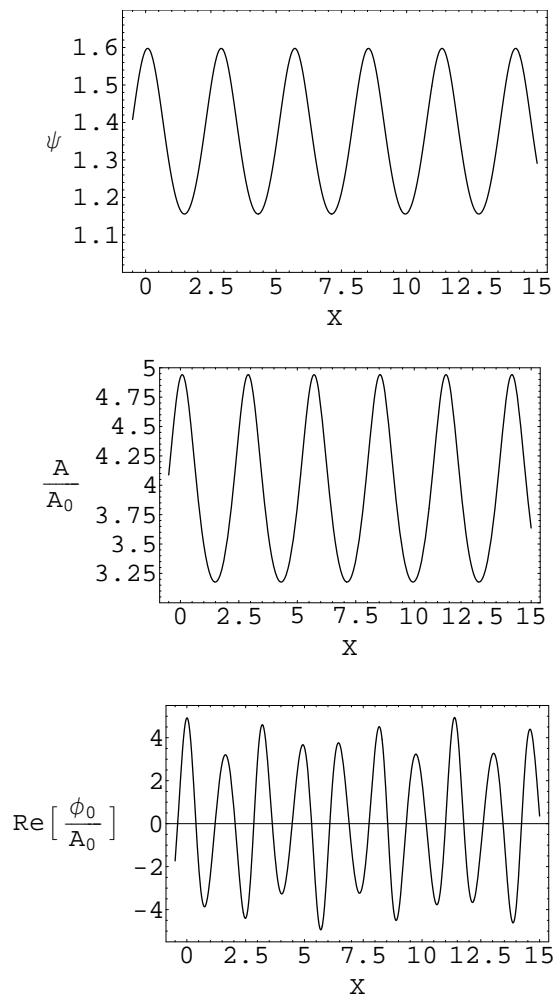}
\caption{Auxiliary function $\psi$ (on the top), amplitude $A$ (on the mid) and real part of the envelope $\phi_0$ (on the bottom) for linear oscillations near the fixed point on the right of the Sagdeev potential. Parameters: $32 (k^2u/\omega)^2 = 1/2, -2\beta\gamma\,k|A_0| = 3, \beta^2 A_{0}^2/2 = 1/4, \Delta = 0$. Initial conditions: $\psi(0) = 1.4, \psi'(0) = 0.5$. The pseudo-energy is then $H = - 0.7$, near the bottom of the right potential well.}
\label{fig5}
\end{center}
\end{figure}

\begin{figure}
\begin{center}
\includegraphics{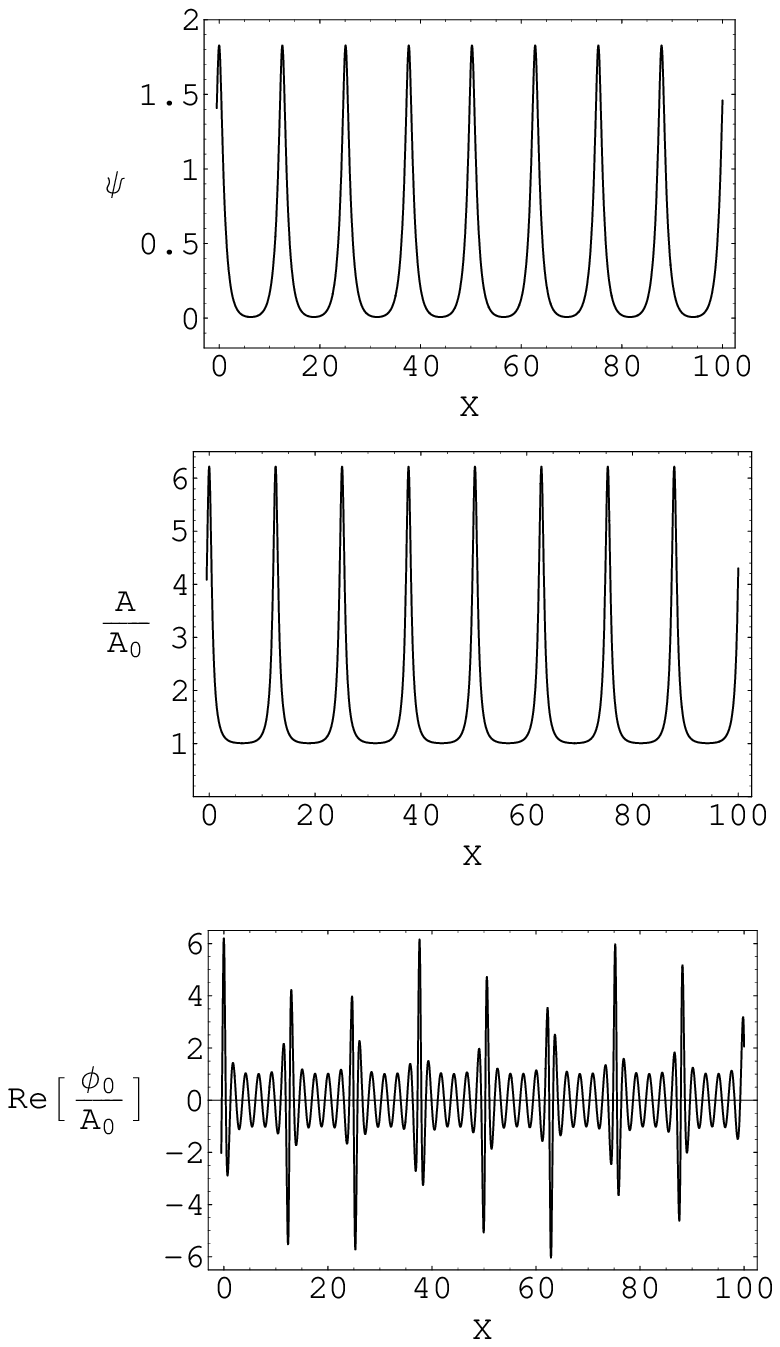}
\caption{Auxiliary function $\psi$ (on the top), amplitude $A$ (on the mid) and real part of the envelope $\phi_0$ (on the bottom) for nonlinear oscillations around the fixed point on the right of the Sagdeev potential. Parameters: $32 (k^2u/\omega)^2 = 1/2, -2\beta\gamma\,k|A_0| = 3, \beta^2 A_{0}^2/2 = 1/4, \Delta = 0$. Initial conditions: $\psi(0) = 1.4, \psi'(0) = 1.3$. The pseudo-energy is then $H = - 10^{-5}$.}
\label{fig6}
\end{center}
\end{figure}

\begin{figure}
\begin{center}
\includegraphics{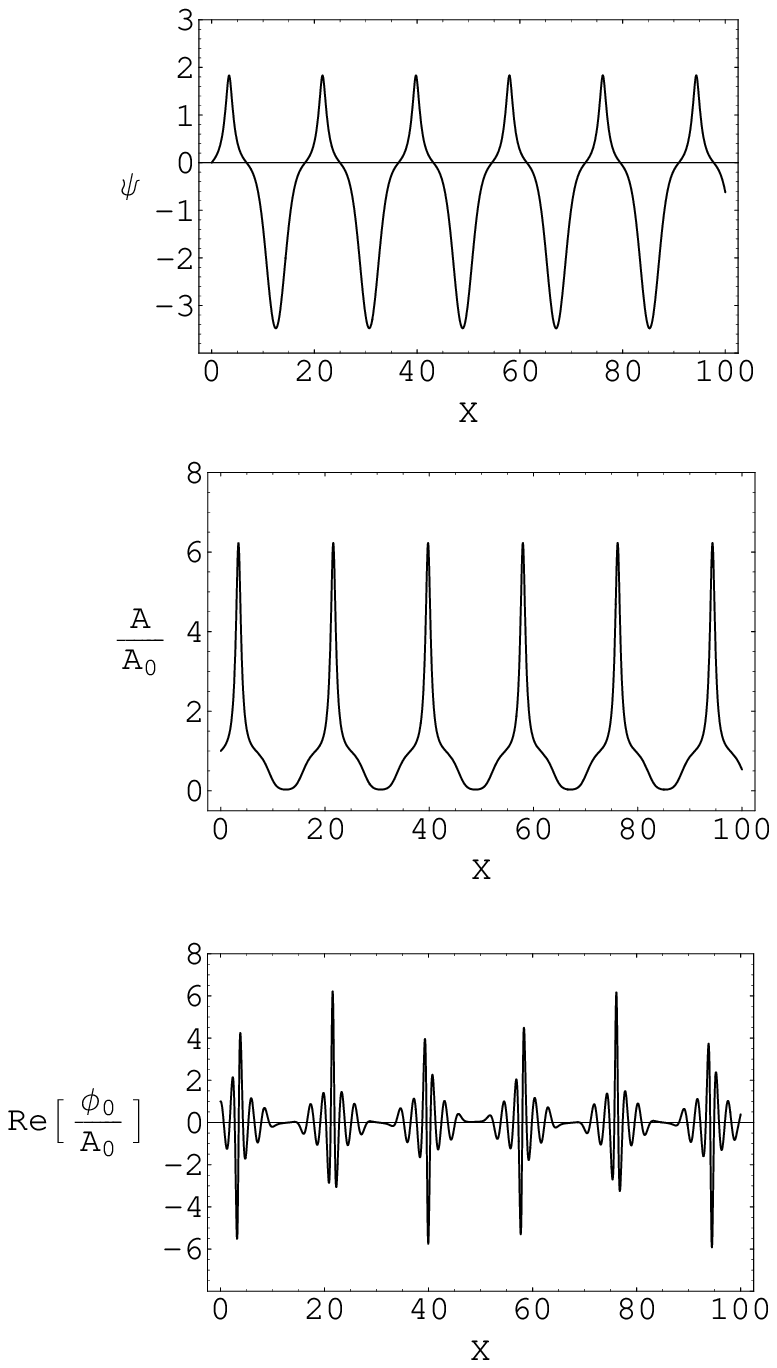}
\caption{Auxiliary function $\psi$ (on the top), amplitude $A$ (on the mid) and real part of the envelope $\phi_0$ (on the bottom) for oscillations around zero, for a slightly positive pseudo-energy. Parameters: $32 (k^2u/\omega)^2 = 1/2, -2\beta\gamma\,k|A_0| = 3, \beta^2 A_{0}^2/2 = 1/4, \Delta = 0$. Initial conditions: $\psi(0) = 0, \psi'(0) = \sqrt{2}/10$. The pseudo-energy is then $H = 10^{-2}$.}
\label{fig7}
\end{center}
\end{figure}

\begin{figure}
\begin{center}
\includegraphics{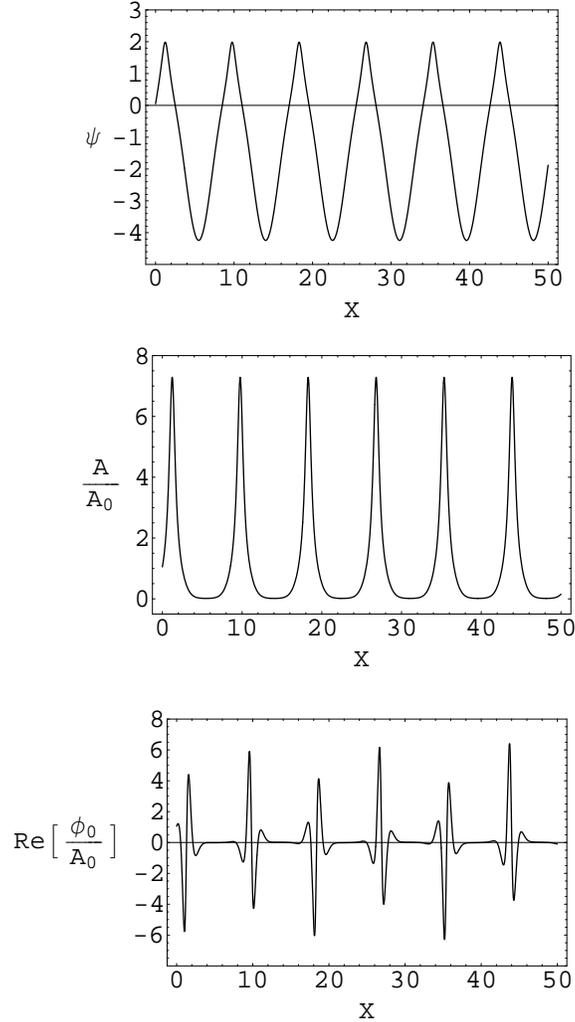}
\caption{Auxiliary function $\psi$ (on the top), amplitude $A$ (on the mid) and real part of the envelope $\phi_0$ (on the bottom) for oscillations around zero, for large pseudo-energy. Parameters: $32 (k^2u/\omega)^2 = 1/2, -2\beta\gamma\,k|A_0| = 3, \beta^2 A_{0}^2/2 = 1/4, \Delta = 0$. Initial conditions: $\psi(0) = 0, \psi'(0) = \sqrt{2}$. The pseudo-energy is then $H = 1$.}
\label{fig8}
\end{center}
\end{figure}

Although the problem was reduced to a quadrature, it can not be done in terms of elementary functions, due to the form of the Sagdeev potential. Nevertheless, the trajectories can be numerically calculated, as shown in Figs. (\ref{fig3}) to (\ref{fig8}), containing the auxiliary function $\psi$, the amplitude $A$ and the real part of the envelope $\phi_0$. In the later, one can see the rapid oscillations due to the incorporation of the phase given by Eq. (\ref{e11}). Figs. (\ref{fig3}-\ref{fig8}) were build for a fixed bi-stable case but varying pseudo-energy $H$ and initial position, so that $\psi$ oscillates around the left equilibrium point, the right equilibrium point or zero (for positive pseudo-energy). A significant distortion takes place according to the degree of nonlinearity. In particular, for positive pseudo-energy, the trajectories of the auxiliary function $\psi$ are not restricted to the left or right equilibria, and nonlinear oscillations around the origin take place, see Figs. (\ref{fig7}-\ref{fig8}).

\section{A conservation law for the Gradov-Stenflo equation and modulational instability}
A peculiar property of Eq. (\ref{e1}) is the singular character of its third term when the envelope $\phi_0$ approaches zero at some point, except in the non-generic situation where the derivative is also zero at that point. The existence of a similar singularity was even an argument to discard a dissipative nonlinear quantum mechanics model introduced by Kibble \cite{Kibble}. Actually, it is apparent from the original derivation \cite{Stenflo, Gradov} that a formal expansion in powers of $k^{-1}\partial\ln\phi_{0}/\partial x$ was used. Clearly such a formal expansion is not valid near a singularity. Therefore, the whole theory becomes invalid in such a case.

In this context, it is natural to assume an slowly varying and everywhere nonzero initial condition, $\phi_{0}(x,0) \neq 0,\, |\partial_{x}\phi_{0}(x,0)/\phi_{0}(x,0)| \ll k, \forall x$. A obvious question is then: is the nonzero property of the envelope field preserved by Eq. (\ref{e1})? Fortunately, the answer is yes, due to the following argument. 

By inspection of Eq. (\ref{e6}), a conservation law is easily detected, 
\begin{equation}
\frac{d}{dt}\,\int\,dx\,\ln\,A = \int\,dx\,\frac{\partial_{t}A}{A} = - \frac{\omega}{8k^2}\,\int\,dx\,\partial_{x}^{2}S = 0 \,,
\end{equation}
for {\it e.g.} decaying or periodic boundary conditions. Therefore, we have a constraint preventing the singularity, provided the initial envelope is everywhere nonzero. 

Also due to the nonzero property of the envelope, it is convenient to linearize Eq. (\ref{e1}) around the exact homogeneous solution
\begin{equation}
\label{e17}
\phi_0 = \bar\phi_0 \exp\left[\frac{i}{8k^2}(\beta^2 |\bar\phi_0|^2 + 2\,\beta\gamma\,k\,|\bar\phi_0|)\,\omega\,t\right] \,,
\end{equation}
representing a plane wave whose frequency is nonlinearly shifted, where $\bar\phi_0$ is a positive amplitude. For simplicity, the case of zero frequency mismatch is being considered ($\Delta \equiv 0$). 

Assuming a perturbation in the form 
\begin{equation}
\label{e18}
\phi_0 = \bar\phi_0\, (1+\tilde{\phi}) \exp\left[\frac{i}{8k^2}(\beta^2 |\bar\phi_0|^2 + 2\,\beta\gamma\,k\,|\bar\phi_0|)\,\omega\,t + i\tilde{\theta}\right] \,,
\end{equation}
where $\tilde{\phi}$ and $\tilde{\theta}$ are real first-order quantities proportional to $\exp(i\tilde{k}x+\sigma\,t)$, we derive the dispersion relation
\begin{equation}
\label{e19}
\sigma^2 = \left(\frac{\omega\,\tilde{k}}{8\,k^2}\right)^2\,\left[2\beta\gamma\,k|\bar\phi_0| + 2\,\beta^2 |\bar\phi_0|^2 - {\tilde{k}}^2\right] \,.
\end{equation}
Notice that the potentially singular term of Eq. (\ref{e1}) has no r\^ole when linearizing around a nonzero solution. In this context this term represents a higher-order nonlinearity.

From Eq. (\ref{e19}), it follows that $\sigma$ is imaginary and the wave amplitude remains bounded when $|\bar\phi_0| < - \gamma\,k/\beta > 0$, which is possible for $k > 0$ since $\gamma/\beta < 0$. 

On the other hand, $\sigma^2 > 0$ when $|\bar\phi_0| \geq - \gamma\,k/\beta$ and $2\beta\gamma\,k|\bar\phi_0| + 2\,\beta^2 |\bar\phi_0|^2 > {\tilde{k}}^2$, so that the amplitude is exponentially amplified. This is a modified modulational instability admitted by Eq. (\ref{e1}).

\section{Conclusion}
As one of our final remarks, we observe that apparently there is no model for the parameter $\gamma$ in the literature, in terms of physical quantities. At the present stage, it should be considered as a phenomenological constant, introduced as a result of a Taylor expansion, see Eq. (\ref{e4}). Also the velocity $u$ can be arbitrarily chosen. Other improvements could be the consideration of moving or diffuse \cite{Brodin} boundaries.  As discussed in Section IV, the model Eq. (1) was derived on the assumption of a slowly varying nonzero envelope. It would be an relevant improvement, to derive the corresponding evolution equation allowing for zero amplitude solutions. These open questions certainly deserve more attention. 

\vskip .5cm
\noindent
{\bf Acknowledgments}
\vskip .5cm
\noindent
F.H. acknowledges the support provided by a fellowship of the Alexander von Humboldt Foundation.

\end{document}